\documentclass[superscriptaddress,twocolumn]{revtex4-1}
\usepackage{amsmath}
\usepackage{color}
\usepackage{amsmath}      	
\usepackage{amssymb} 
\usepackage{amsfonts}   
\usepackage{graphicx}
\usepackage{bm}
\usepackage{multirow}
\usepackage{nicefrac}
\usepackage{url}

\newcommand{\LPTHE}{Laboratoire de physique th\'eorique et hautes \'energies, CNRS and Sorbonne Universit\'e, 4 Place Jussieu, 75005 Paris, France}
\newcommand{\LPS}{Laboratoire de physique statistique, CNRS, Sorbonne Universit\'e, Universit\'e Paris-Diderot, and \'Ecole normale sup\'erieure (PSL), 24 rue Lhomond, 75005 Paris, France}
\newcommand{\LPT}{Laboratoire de physique th\'eorique, CNRS, Sorbonne Universit\'e, and \'Ecole normale sup\'erieure (PSL), 24 rue Lhomond, 75005 Paris, France}
\newcommand{\lastequal}{These authors contributed equally. Please correspondance to \url{tmora@lps.ens.fr}, \url{awalczak@lpt.ens.fr}}

\makeatletter
\def\@seccntformat#1{%
  \expandafter\ifx\csname c@#1\endcsname\c@section\else
  \csname the#1\endcsname\quad
  \fi}
\makeatother

\begin{document}

\title{Genesis of the $\alpha\beta$ T-cell receptor}

\author{Thomas Dupic}
\affiliation{\LPTHE}
\author{Quentin Marcou}
\affiliation{\LPT}
\author{Aleksandra M. Walczak}
\thanks{\lastequal}
\affiliation{\LPT}
\author{Thierry Mora}
\thanks{\lastequal}
\affiliation{\LPS}

\begin{abstract}
The T-cell (TCR) repertoire relies on the diversity of receptors composed of two chains, called $\alpha$ and $\beta$, to recognize pathogens. Using results of high throughput sequencing and computational chain-pairing experiments of human TCR repertoires, we quantitively characterize the $\alpha\beta$ generation process. We estimate the probabilities of a rescue recombination of the $\beta$ chain on the second chromosome upon failure or success on the first chromosome. Unlike $\beta$ chains, $\alpha$ chains recombine simultaneously on both chromosomes, resulting in correlated statistics of the two genes which we predict using a mechanistic model. We find that $\sim 28 \%$ of cells express both $\alpha$ chains. 
Altogether, our statistical analysis gives a complete quantitative mechanistic picture that results in the observed correlations in the generative process. We learn that the probability to generate any TCR$\alpha\beta$ is lower than $10^{-12}$ and estimate the generation diversity and sharing properties of the $\alpha\beta$ TCR repertoire.

\end{abstract}

\maketitle

\section*{Introduction}

The adaptive immune system confers protection against many different pathogens using a diverse set of specialized receptors expressed on the surface of T-cells. The ensemble of the expressed receptors is called a repertoire and its diversity and composition encode the ability of the immune system  to recognize antigens. T-cell receptors (TCR) are composed of two chains, $\alpha$ and $\beta$, that together bind antigenic peptides presented on the multihistocompatability complex (MHC). High-throughput immune sequencing experiments give us insight into the repertoire composition through lists of TCR, typically centered around the most diverse region, the Complimentary Determining Region 3 (CDR3) of these chains \cite{Robins2009,Boyd2009a,Benichou2012,Robins2013a,Six2013}. Until recently most experiments and analyses focused on only one of the two chains at a time, and studies of TCR with both chains were limited to low-throughput methods \cite{Kim2012,Turchaninova2013c,Han2014}.
Recent technological and analytical breakthroughs now allow us to simultaneously determine the sequences of both $\alpha$ and $\beta$ chains expressed on cells of the same clone in a high-throughput way \cite{HowieHighthroughputpairingcell2015} (see also analysis of unpublished data obtained by single-cell sequencing in \cite{Grigaityte2017}). These advances make it possible to study the repertoires of paired receptors, and to revisit the questions of the generation, distribution, diversity and overlap of TCR repertoires previously studied at the single-chain level \cite{Robins2010,Murugan2012,Qi2014,Elhanati2014,Mora2016e,Pogorelyy2017,ElhanatiPredictingspectrumTCR2018}, but also to gain insight into the mechanisms of T-cell recombination and maturation.

\begin{figure}[!h]
  \begin{center}
\includegraphics[width=\linewidth]{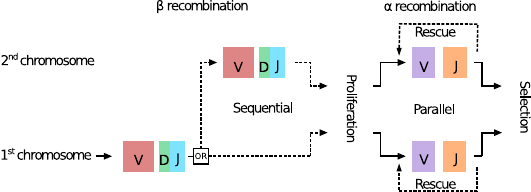}
\caption{{\bf Formation of a T-cell receptor.} The $\beta$ chain is rearranged before the $\alpha$ chain. The recombination on the two chromosomes is sequential for $\beta$, and parallel for $\alpha$. Dotted lines indicate optional events. Rescue events on the $\alpha$ chain correspond to successive recombinations of the same locus (see also schematic in Fig.~\ref{fig:pair_correlation}).
\label{fig:tree_proba}
}
\end{center}\end{figure}

TCR receptor diversity arises from genetic recombination of the $\alpha$ and $\beta$ chains of thymocytes in the thymus.
Each chain locus consists of a constant region (C), and multiple gene segments V ($52$ for the human $\beta$ chain and $\approx\ 70$ for $\alpha$), D ($2$ and $0$) and J ($13$ and $61$). Recombination proceeds by selecting one of each type of segment and joining them together, with additional deletions or insertions of base pairs at the junctions. TCR$\beta$ is first recombined and expressed along with the pre-T cell receptor alpha (a non-recombined template gene) on the surface of the cell to be checked for function. T cells then divide a few times before TCR$\alpha$ recombination begins, at which point the thymic selection process acts on the complete receptor. The recombination of each chain often result in non-productive genes (e.g. with frameshifts or stop codons). Subsequent rescue and selection mechanisms ensure that all mature T cells express at least one functional receptor. Recombination of the $\beta$ chain on the second chromosome may be attempted if the initial recombination was unsuccessful. By contrast, the $\alpha$ chain is recombined on both chromosomes simultaneously \cite{PetrieMultiplerearrangementscell1993}, and proceeds through several recombination attempts that successively join increasingly distal V and J segments (Fig.\,\ref{fig:tree_proba}). Taken together, recombination events can potentially produce up to 4 chains (2 $\alpha$ and 2 $\beta$) in each cell. In principle, allelic exclusion ensures that only one receptor may be expressed on the surface of the cell, but this process is leaky: $7\%$ of T-cells have two productive $\beta$-chains \cite{Stubbingtoncellfateclonality2016, EltahlaLinkingcellreceptor2016}, and $\%1$ express both of them on the surface  \cite{DavodeauDualcellreceptor1995, PadovanNormallymphocytescan1995, SteinelPostTranscriptionalSilencingVvDJvCv2010}. Allelic exclusion in the $\alpha$ chain is less well quantified as it relies on different mechanisms \cite{RybakinAllelicExclusionTCR2014, NiederbergerAllelicExclusionTCR2003}, with estimates ranging from 7\% \cite{Han2014} to 30\% \cite{PadovanNormallymphocytescan1995} of cells with two functionally expressed $\alpha$ chains.

Despite the partial characterization of the various mechanisms underpinning the recombination, rescue and selection of the two TCR chains, a complete quantitative picture of these processes is still lacking. For instance, the probability of recombination rescue, the probability for a chain to pass selection, or the extent of allelic exclusion, have not been measured precisely.
Here we re-analyse the data from \cite{HowieHighthroughputpairingcell2015} to link together each of the 4 $\alpha$ and $\beta$ chains of single clones, and study $\alpha$-$\alpha$ and $\beta$-$\beta$ pairs as well as $\alpha$-$\beta$ pairs. Using these pairings, we propose a mechanistic model of recombination of the two chains on the two chromosomes, inspired by \cite{WarmflashModelTCRGene2006}, and study the statistics of the resulting functional $\alpha\beta$ TCR.

\section*{Results}

\subsection*{Pairing multiple chains in the same clone}

We analysed previously published data on sequenced T-cell CDR3 regions obtained from two human subjects (PairSEQ), as described by Howie and collaborators \cite{HowieHighthroughputpairingcell2015}. In the original study, sequences of  $\alpha$ and $\beta$ chain pairs associated to the same clone were isolated using a combination of high-throughput sequencing and combinatorial statistics.
Briefly, T cell samples were deposited into wells of a 96-well plate, their RNA extracted, reverse-transcribed into cDNA with the addition of a well-specific barcode, amplified by PCR, and sequenced. $\alpha\beta$ pairs appearing together in many wells were assumed to be associated with the same T-cell clone, and thus expressed together in the same cells.
Because the method relies on the presence of cells of the same clone in many wells, the method can only capture large memory T cell clones present in multiple copies in the same blood sample. Naive clones which have a population size of around $10$, or concentration of $10^{-10}$ \cite{CasrougeSizeEstimateav2000}, are not expected to be paired in this way.

We generalized the statistical method of \cite{HowieHighthroughputpairingcell2015} to associate $\alpha$-$\alpha$ and $\beta$-$\beta$ pairs present in the same clone. Along with $\alpha$-$\beta$ pairings, this allowed us to reconstruct the full TCR content of a cell. Two additional difficulties arise when trying to pair chains of the same type. First, truly distinct pairs of chains must be distinguished from reads associated with the same sequence but differing by a few nucleotides as a result of sequencing errors.
We set a threshold of $11$ nucleotide mismatches on the distribution
of distances between paired chains (Fig.\,S1) to remove
  duplicates while minimizing the loss of real pairs.
Second, because of allelic exclusions, one of the two chains of the same type is typically expressed in much smaller amounts than the other. As a result, we find much fewer $\alpha$-$\alpha$ and $\beta$-$\beta$ pairs than $\alpha$-$\beta$ pairs.

Table \ref{tab:number_pairs} summarizes the numbers of pairs found in each experiment, with a significance threshold chosen to achieve a 1\% false discovery rate (see Methods). This method can then be used to recreate the complete TCR content of a given clone, and set apart clones expressing multiple TCR receptors.

\subsection*{Correlations between chains of the same cell}
\begin{figure}[!h]
  \begin{center}
\includegraphics[width=\linewidth]{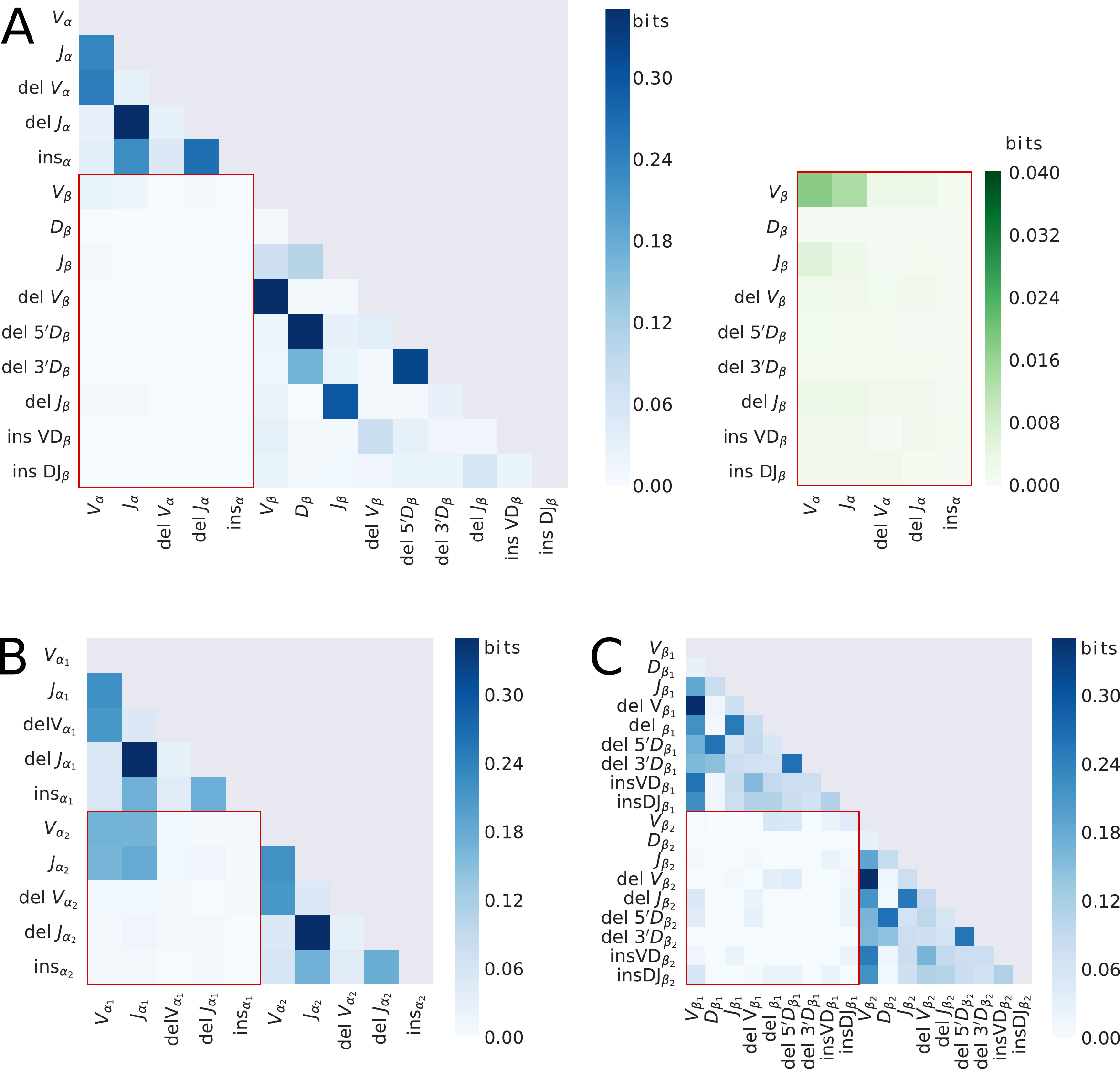}
\caption{ {\bf Mutual information (a non-parametric measure of correlations) between the recombination events of the paired chains:} V, D, and J segment choice, numbers of bases deleted from the 3' end of the V-gene (delV), the $5^{\prime}$ end of the J-gene (delJ), and both ends of the D-gene for the $\beta$ chain (del$5^\prime$D and del$3^\prime$D for the $5^\prime$ and $3^\prime$ ends, respectively); number of insertions of random nucleotides between V and J segments (insVJ) for the $\alpha$ chain, and between V and D (insVD) and between J and D (insDJ) segments for the $\beta$ chain. Mutual information for ({\bf A}) $\alpha$-$\beta$ pairs (on the right in green: close-up of the inter-chain mutual information); ({\bf B}) $\alpha$-$\alpha$ pairs; and ({\bf C}) $\beta$-$\beta$ pairs. Inter-chain correlations are highlighted by red boxes. To remove systematic biases in mutual information estimation from finite data, the mutual information of shuffled data was subtracted (see Methods). For a statistical anlaysis of the significance of the reported mutual informations, see Fig.\,S2.
\label{fig:pair_mutual_information}
  }
\end{center}\end{figure}

Correlations between the features of the recombination events of the chains present in the same cells are informative about the rules governing the formation of a mature $\alpha\beta$ TCR in the case of $\alpha$-$\beta$ pairings, and also about the mechanisms and temporal organization of recombination on the two chromosomes in the case of $\alpha$-$\alpha$ and $\beta$-$\beta$ pairings. We computed the mutual information, a non-parametric measure of correlations (see Methods), between pairs of recombination features for each chain: V, D, and J segment choices, and the numbers of deletions and insertions at each junction (Fig.\,\ref{fig:pair_mutual_information}).
Because recombination events cannot be assigned with certainty to a given sequence, we used the IGoR software \cite{MarcouHighthroughputimmunerepertoire2018} to associate recombination events to each sequence with a probabilistic weight reflecting the confidence we have in this assignment (see Methods). We have shown previously that this probabilistic correction removes spurious correlations between recombination events \cite{Murugan2012,MarcouHighthroughputimmunerepertoire2018}. 
Correlations within single chains recapitulate previously reported results for the $\beta$ \cite{Murugan2012} and $\alpha$ \cite{Elhanati2016a} chains. Inter-chain correlations, highlighted by red boxes, are only accessible thanks to the chain pairings.

We find no correlation between the number of insertions in different chains across all pair types. Such a correlation could have been expected because Terminal deoxynucleotidyl transferase (TdT), the enzyme responsible for insertions, is believed to correlate with the number of inserted base pairs \cite{MoteaTerminalDeoxynucleotidylTransferase2010}, and is expected to be constant across recombination events in each cell. The lack of correlation between different insertion events thus suggests that the there is no shared variability arising from differences in TdT concentration across cells.

We report generally weak correlations between the $\alpha$ and $\beta$ chains (Fig.\,\ref{fig:pair_mutual_information}A and Fig.\,S2 for an analysis of statistical significance), with a total sum of $0.36$ bits, about 10 times lower than the total intra-gene correlations of the $\alpha$ chain. The largest correlation is between the choice of $V_\alpha$ and $V_\beta$ genes ($0.036$ bits) and $J_\alpha$ and $V_\beta$ genes ($0.033$ bits), in agreement with the analysis of \cite{Grigaityte2017} on unpublished single-cell data.
These correlations probably do not arise from biases in the recombination process, because recombination of the two chains occurs on different loci (located on distinct chromosomes) and at different stages of T cell maturation. A more plausible explanation is that thymic selection preferentially selects some chain associations with higher folding stability or better peptide-MHC recognition properties. Distinguishing recombination- from selection-induced correlations would require analysing pairs of non-productive sequences, which are not subjected to selection, but the number of such pairs in the dataset was too small to extract statistically significant results. An analysis of the correlations between gene segments (Fig.\,S3) does not show any particular structure.

Pairs of $\beta$ chains show almost no correlations (Fig.\,\ref{fig:pair_mutual_information}C and Fig.\, S2 for an analysis of statistical significance).
Looking in detail at the correlations between gene segments reveals a strongly negative correlation of TCRBV21-01 and TCRBV23-01 (both pseudogenes) with themselves (Fig.\,S4), which is expected because at least one of the two $\beta$ chain must have a non-pseudogene V. More generally, correlations are likely to arise from selection effects, since the two recombination events of the two $\beta$ chains are believed to happen sequentially and independently. The fact that at least one of the chains needs to be functional for the cell to survive breaks the independence between the two recombination events.

By contrast, the $\alpha$-$\alpha$ pairs have very strong correlations between the V and J usages of the two chromosomes, and none between any other pair of features  (Fig.\,\ref{fig:pair_mutual_information}B). These correlations arise from the fact that the two $\alpha$ recombination events occur processively and simultaneously on the two chromosomes, as we analyse in more detail below.

\subsection*{Correlations between $\alpha$ chains can be explained by a rescue mechanism}

\begin{figure}[!h]
  \begin{center}
\includegraphics[width=\linewidth]{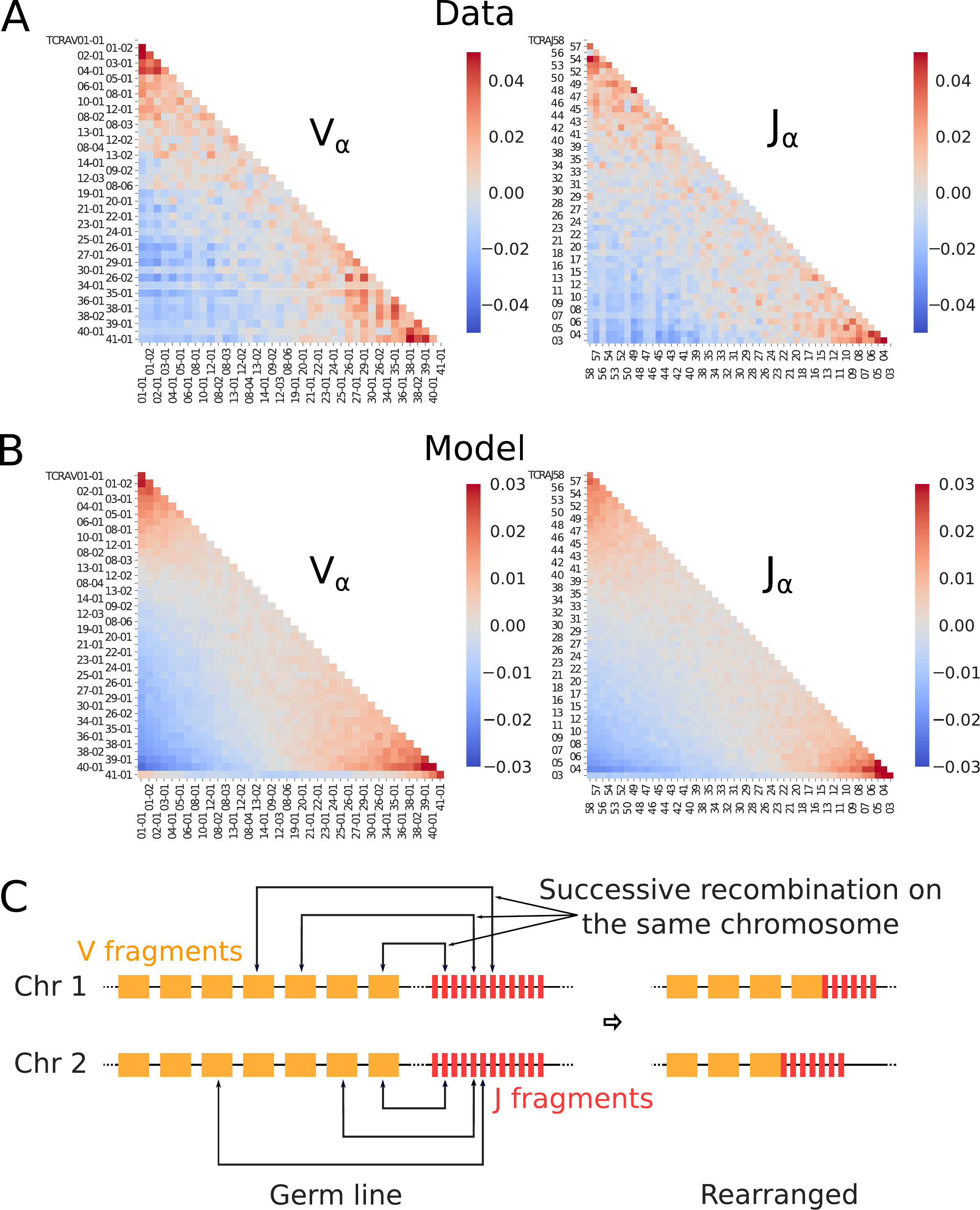}
\caption{{\bf Evidence of the rescue mechanism. }
  (A) Pearson correlation between V and J gene segment usage for TCR$\alpha$. The correlation is taken between the truth values of particular V and J gene choices (a value of 1 is assigned if a given segment is observed and 0 if it is not, see Methods for details).
  (B) Same Pearson correlation as in (A) calculated from simulations of the rescue mechanism model depicted in (C).
  (C) Cartoon of the rescue mechanism. The rescue happens simultaneously on the two chromosomes. Once one of the re-arrangements results in a functional rearrangement, recombination stops. In the end, the V and J gene segments selected on both chromosomes are close to each other in the germline ordering.
\label{fig:pair_correlation}
}
\end{center}\end{figure}

We wondered whether the detailed structure of the observed correlations between the $\alpha$ chains on the two chromosomes could be explained by a simple model of recombination rescue.
The correlations of the $V_{\alpha}$ segments on the two chromosomes and of the $J_{\alpha}$ segments show a similar spatial structure as a function of their ordering on the chromosome (see Fig~\ref{fig:pair_correlation}A): proximal genes are preferentially chosen together on the two chromosomes, as are distal genes. The correlations between the $V_\alpha$ gene segment on the first chromosome and the $J_\alpha$ on the second chromosome also show a similar diagonal structure (Fig.\,S5).

The two chromosomes recombine simultaneously, and proceed by successive trials and rescues. If the first recombination attempt fails to produce a functional chain, another recombination event may happen on the same chromosome between the remaining distal V and J segments, excising the failed rearranged gene in the process.
The recombination of a functional chain on either of the chromosomes immediately stops the process on both chromosomes.
By the time this happens on one chromosome, a similar number of recombination attempts will have occurred on the other chromosome. We hypothesize that this synchrony is the main source of correlations between the V$\alpha$ and J$\alpha$ gene usages of the two chains.

To validate this hypothesis, we simulated a minimal model of the
rescue process similar to \cite{WarmflashModelTCRGene2006} (Methods),
in which the two chromosomes are recombined in parallel. If
recombination happens to fail on both chromosomes, repeated ``rescue''
recombinations (which we limit to 5) take place between outward nearby segments (Fig~\ref{fig:pair_correlation}C).
The covariance matrices obtained from the simulations for both $V_{\alpha}$ and $J_{\alpha}$ (Fig.\,\ref{fig:pair_correlation}B) show profiles that are very similar to the data, with positive correlations along the diagonal, in particular at the two ends of the sequence. However, the actual distributions of V and J genes segments (see Fig.\,S6) are much more heterogeneous than the slowly decaying distribution implied by our simple model: the question of gene usage is further complicated by other factors, such as gene accessibility and primer specificity.

\subsection*{Probability of recombination of the second chromosome} 

We wondered if the paired data could be used to estimate the percentage of cells with two recombined chains of the same type. However, since pairing was done based on mRNA transcripts through cDNA sequencing, silenced or suppressed genes are not expected to be among the identified pairs, leading to a systematic underestimation of double recombinations. While the authors of \cite{HowieHighthroughputpairingcell2015} also provided a genomic DNA (gDNA) dataset that does not have this issue, the number of sequences was too small to resolve statistically significant pairings.
Nonetheless, we can derive strict bounds from the proportion of productive sequences found in this (unpaired) gDNA dataset. 
Following recombination, using IGoR we estimate $p^\alpha_{\text{nc}} = 69.5 \%$ of the $\alpha$ sequences, and $p^\beta_{\text{nc}} = 73.5 \%$ of $\beta$ sequences are non-coding or contain a stop codon. We collectively refer to as ``non-coding'' sequences. The remaining sequences, called ``coding'', make up a fraction $p^{\alpha,\beta}_{\rm c}=1-p^{\alpha,\beta}_{\rm nc}$ of random rearrangements.
We denote by $p^\alpha_{\rm f}$ and $p^\beta_{\rm f}$ the probability that a coding sequence can express a functional $\alpha$ or $\beta$ chain that can ensure its selection.

\begin{figure}[!h]
  \begin{center}
\includegraphics[width=\linewidth]{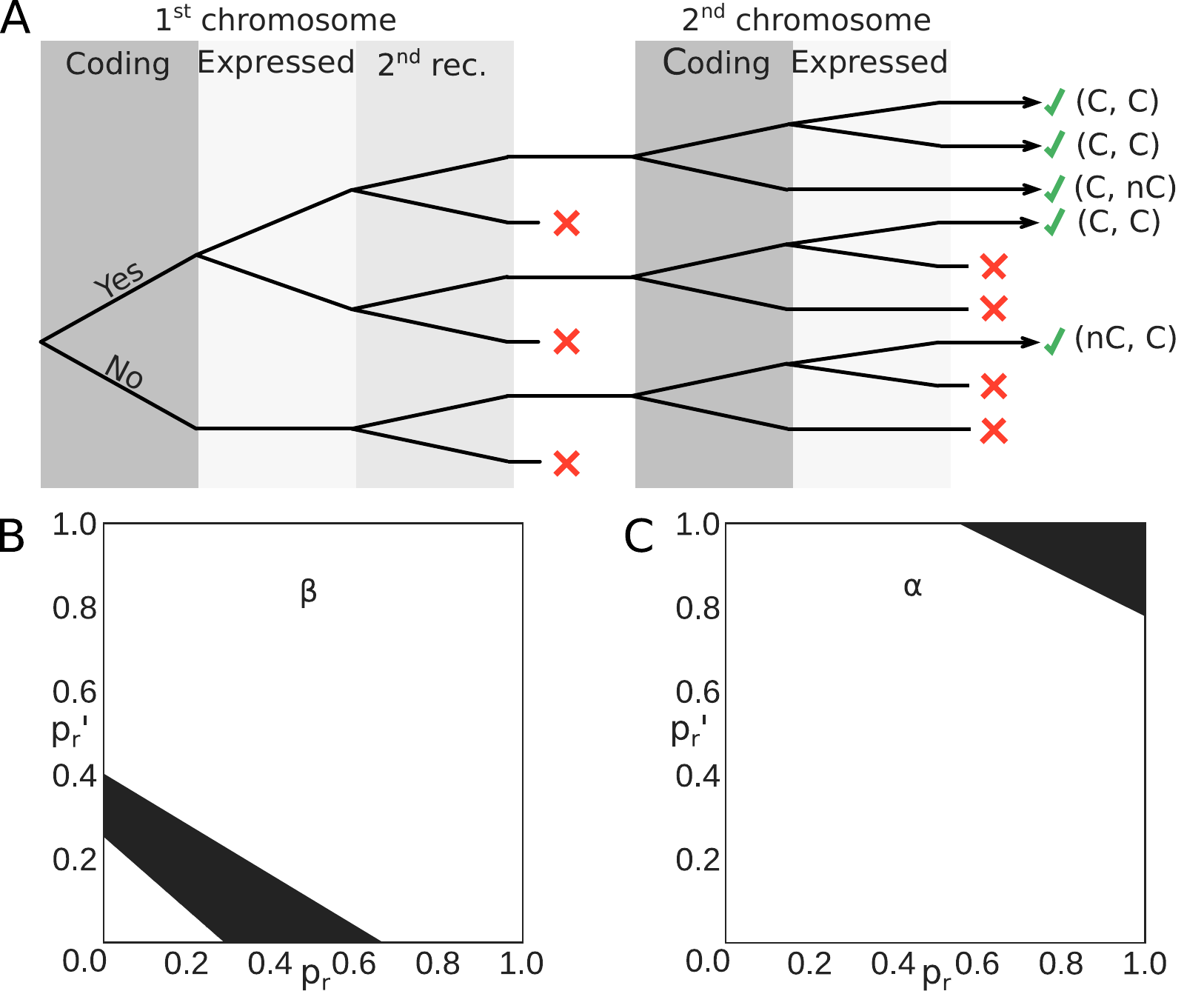}
  \caption{{\bf Probability of recombination of the second chromosome}
    (A) Decision tree of the recombination process for one chain ($\alpha$ or $\beta$). The first part shows the recombination of the first chromosome, the second part of the second chromosome. In each area a binary choice is made. Red crosses indicate decision outcomes that lead to no observed sequence. Observable outcomes (with at least one coding sequence) are indicated at the end of the tree by green ticks. C stands for coding, nC for non-coding. 
  (B) Bounds on the allowed values of rescue probabilities for the $\beta$ chain calculated from the decision tree in (A). The black part of the graph corresponds to the allowed values of $p_r'$ (probability of a second recombination for $\beta$ if the first was successful) and $p_r$ (probability of a second recombination for $\beta$ if the first was not successful). The bounds were obtained by imposing $0 < p_f^{\beta} < 1$ in Eq.\,\ref{eq:proportion_beta}.
   (C) Bounds on the allowed values of rescue probabilities for the $\alpha$ chain. They are consistent with both chromosomes recombining simultaneously and independently, $p_r=p_r'=1$.
\label{fig:beta_allowed_proba}
}
\end{center}\end{figure}

The number of observed non-coding sequences depends on whether the second chromosome attempts to recombine following the recombination of the first one. We call $p_{r}$ the probability that a second recombination happens when the first recombination fails to produce a functional chain, and $p_{\rm r}'$ when the first recombination succeeds.
Then, the proportion $f_{\rm nc}$ of observed non-coding sequences can be written as (see tree in Fig.\,\ref{fig:beta_allowed_proba} and Methods):
\begin{equation}
f_{\rm nc}=  \frac{(p_{\rm r}+p_{\rm r}')p_{\text{nc}}}{1 + p_{\rm r}' + 2(1-p_{\rm f}p_{\rm c})p_{\rm r}}.
  \label{eq:proportion_beta}
\end{equation}
Note that this formula assumes that the presence of more than one functional chain does not affect its selection probability.
Comparing the proportion of observed non-coding $\beta$ chain sequences calculated from Eq.\,\ref{eq:proportion_beta} with the values from gDNA data ($f_{\rm nc}^\beta= 18 \pm 1 \%$ in \cite{HowieHighthroughputpairingcell2015} and 14\% in \cite{Robins2010}), 
allows us to constrain the values of $p^\beta_{\rm r}$ and ${p^{\beta}_{\rm r}}'$. The probability of a second recombination, even if the first recombination failed, is always lower than $65\%$ (Fig~\ref{fig:beta_allowed_proba}A).
By constrast, the observed fraction of non-coding sequences in the $\alpha$ chain, $f_{\rm nc}^\alpha=40\pm 1 \%$, constrains the the rescue probabilities $p^\alpha_{\rm r}$ and ${p^\alpha_{\rm r}}'$ to be close to 100\% (Fig \ref{fig:beta_allowed_proba}B), in agreement with the fact that both chromosomes are believed to recombine independenly.
Assuming strict independence, $p^\alpha_{\rm r}={p^\alpha_{\rm r}}'=1$ puts bounds on the probability that a random coding $\alpha$ sequence is functional, $70\%\leq p^\alpha_{\rm f}\leq 100\%$.

\subsection*{Fraction of cells with two functional $\alpha$ chains} 

Can we learn from pairing data what fraction of cells expressed two chains of the same type? gDNA pairings do not allow us to do that, because they are severely limited by sequencing depth: most chains cannot be paired because of material losses, and estimating the fraction of cells with several chains is impossible. While cDNA pairings are in principle less susceptible to material loss, non-functional sequences are much less expressed than functional ones \cite{NiederbergerAllelicExclusionTCR2003, SteinelPostTranscriptionalSilencingVvDJvCv2010}, lowering their probability of being paired and introducing uncontrolled biases in the estimate of fractions of cells with different chain compositions.
However, we can use this difference in expression patterns by
examining the distribution of read counts for each type of
chain. For each sequence, we sum the number of its reads in
  all wells. Only the third experiment (the most data rich) is considered to avoid pooling together datasets with different sequencing depths.
Sequences of chains paired with a non-coding chain of the same type must be functional and expressed on the surface of the cell. Those sequences have a markedly different distribution of read counts than non-coding sequences (Fig.\,S7A and B). Coding sequences that are coupled with another coding sequence can be either expressed or silenced, depending on their own functionality and the status of the other chain. Thus, their read count should follow a mixture distribution of both expressed and silenced sequences, the latter being assumed to follow the same distribution as noncoding sequences. 
Fitting the parameters of this mixture to the read counts of paired
coding sequences (Fig.\,S7C) yields the total proportion $p_{\rm
  e}\leq p_{\rm f}$ of functional and expressed sequences, among all
functional coding sequences $ p_{\rm f}$. The mixture is
  expected to better represent the distribution at high read counts,
  where drop-out effects are less likely to lead to loss of pairs.

For $\alpha$ sequences, we found $p^\alpha_{\rm e}=64\pm 5\%$, meaning that $2p_{\rm e}^\alpha-1=28 \% \pm 10 \%$ of cells express two different $\alpha$ chains (see Methods). 
This number is consistent with older results
\cite{PadovanExpressiontwocell1993}, but slightly higher than a recent
estimate of $14\%$ based on single-cell sequencing
\cite{Han2014}. However, that estimate may be affected by
  material loss and should be viewed as a lower bound. Another
  estimate from the same data \cite{PadovanExpressiontwocell1993}, but
  taking into account material loss (see Methods), suggests that
  $24\pm 5\%$ of cells have two functional and expressed $\alpha$ chains, consistent with our own estimate.

For $\beta$ chains, the fit is noisier, because non-coding sequences are much more suppressed and therefore scarcer than for the $\alpha$ chain (only $4.5$\% of sequences are non-coding).
We estimate that there are 8-10 times more silenced coding sequences than non-coding sequences, but the fit does not allow us to estimate the fraction of cells with two expressed $\beta$ chains, although this number is consistent with 0 according to the data.

\subsection*{Functional sequences are more restricted than `just coding' sequences}

It is often assumed that all coding sequences must be functional, and previous studies have used the difference between coding and non-coding sequences to quantify the effects of selection \cite{Elhanati2014,Elhanati2015,Toledano2018}. However, some fraction of coding sequences may actually be disfunctional, silenced, or not properly expressed on the cell surface. By contrast, sequences that can be paired with a non-coding sequence of the same type must be functional and expressed on the cell surface, lest the cell that carries them dies. These sequences represent a non-biased sample of all functional sequences, and their statistics may differ from those of `just coding' sequences. In Table \ref{tab:shannon_entropy} we report the differences between the two ensembles in terms of CDR3 length (defined from the conserved cystein of V and the conserved phenylalanine or tryptophan of J, corresponding to IMGT positions 105 to 117) and gene usage. 
All comparisons are with sequences that could be paired with another one to remove possible biases from the pairing process.

We find that functional sequences are on average slightly larger (by 1-2 nucleotides) than coding and non-coding sequences (Table \ref{tab:shannon_entropy} and Fig.\,S8). More markedly, the variance of their length is smaller, implying stronger selection towards a prefered length in the functional ensemble than in the coding and non-coding ensembles. These observations, which hold for both the $\alpha$ or $\beta$ chains, indicate that the functional ensemble (as defined here using pairing information) is more restricted than `just coding' sequences, and gives a more precise picture of the selected repertoire.

The impact of selection can also be measured by how much gene usage departs from the unselected ensemble using the Kullback-Leibler divergence (see Methods and Table \ref{tab:shannon_entropy}), offering a more contrasted view.  $V_{\beta}$ and $J_{\alpha}$ usages are similar in functional and coding sequences in terms of their divergence with non-coding sequences. For $J_{\beta}$ however, this divergence is higher in functional than in simply coding sequences, while the opposite is true for $V_\alpha$.

\subsection*{Model predicts very rare $\alpha\beta$ TCR sharing} 

\begin{figure}[!h]
  \begin{center}
\includegraphics[width=\linewidth]{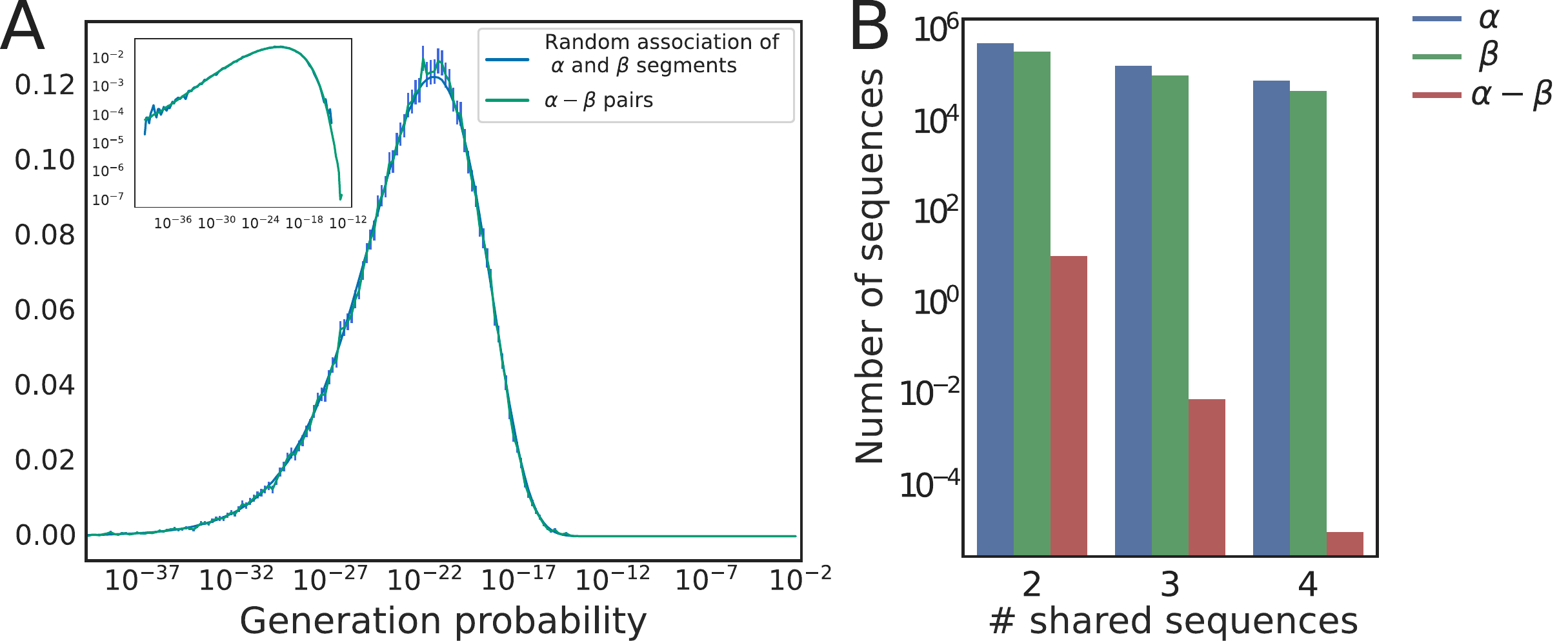}
\caption{{\bf Generation probability of a full $\alpha\beta$ TCR.} (A)
  Distribution of the generation probabilities of $\alpha\beta$ pairs,
  obtained by multiplying the generation probabilities of the $\alpha$
  and $\beta$ sequences. The graph shows the distribution for paired
  sequences (blue) and random associations of $\alpha\beta$ pairs
  (green). The error bars represent three standard deviations,
    and the inset shows the same plot on a double logarithmic scale.
(B) Number of CDR3 nucleotide sequences found in $n$ among 10 individuals with a sample depth of $N = 10^6$ unique $\alpha\beta$ TCR per individual. The probability of more than two people sharing the same TCR receptor is extremely small.
\label{fig:sharing}
}
\end{center}\end{figure}

Ignoring small correlations between features of the $\alpha$ and $\beta$ chains reported in Fig.\,\ref{fig:pair_mutual_information}, we can assume that the probability of generating a $\alpha\beta$ pair is given by the product of the probabilities of generating each chain independently. These probabilities can be calculated using the IGoR software \cite{MarcouHighthroughputimmunerepertoire2018} for each paired chain in our datasets. The distribution of the pair generation probabilities obtained in this way (Fig.\,\ref{fig:sharing}~A) shows an enormous breadth, spanning more than 20 orders of magnitude. We self-consistently validated the assumption of independence by showing that random assortments of $\alpha$ and $\beta$ chains yielded an identical distribution of generation probabilities (green curve).

The maximum TCR generation probability is $< 10^{-12}$, meaning that generating the same pair twice independently is extremely unlikely. This suggests that, without strong antigenic selection, only a negligible number of full TCR sequences will be shared in samples obtained from distinct individuals. To make that prediction more quantitative, we simulated a computational model of sequence generation followed by thymic selection. $\alpha$ and $\beta$ chains were generated by IGoR, and then each TCR$\alpha\beta$ amino-acid sequence was kept with probability $q$ to mimick thymic selection \cite{ElhanatiPredictingspectrumTCR2018}. We further assume that selection acts on each chain independently, so that the ratio $q$ is given by $q_\alpha q_\beta$, where $q_{\alpha,\beta}$ are the selection probabilities infered from the analysis of single chains. These selection factors can be obtained by fitting the curve giving the number of unique amino-acid sequences as a function of unique nucleotide sequences \cite{ElhanatiPredictingspectrumTCR2018}, yielding $q_\beta=0.037$ and $q_{\alpha} = 0.16$ (Fig.\,S9).

Using the model, we can make predictions about the expected number of TCR$\alpha\beta$ nucleotide sequences shared between any of 10 individuals (Fig.\,\ref{fig:sharing}~B) for which a million unique synthetic TCR$\alpha\beta$ were obtained. We find that, while a substantial fraction of sequences of each chain are expected to be shared by several individuals, sharing the full TCR$\alpha\beta$ is very unlikely, and drops well below 1 for more than 2 individuals. This suggests that the existence in real data of any TCR$\alpha\beta$ shared between several individuals should be interpreted as resulting from strong common selection processes, probably associated with antigen-specific proliferation, leading to convergent selection of the shared sequences.
A concomitant question concerns the total number of TCR
  sequences shared between two individuals. This number does not
  depend on selection or sample size, but rather on the total number
  of different clonotypes in an individual. While this last quantity
  is not precisely known, estimates range between $10^8$ and
  $10^{11}$ \cite{Qi2014,Lythe2015}. Using the analytical formulas and numerical procedure
  described in \cite{ElhanatiPredictingspectrumTCR2018} with these
  estimates of the repertoire size, we predict the proportion of shared clonotypes between two individuals to fall between $0.001 \%$ and $0.1\%$ of their full repertoires (see Methods for details).

\subsection*{Co-activation of cells sharing the same $\beta$ chain}
To further investigate the effects of convergent selection, we
quantified how often the same $\alpha$ chain was associated with
distinct $\beta$ chains in different clones (Fig.\,S10A), and {\it
  vice versa}  (Fig.\,S10B). While association of a $\beta$ with 2
distinct $\alpha$ chains could happen in the same cell because of the
existence of two copies, we found a substantial fraction (3\%) of all
paired TCR$\beta$ that could be associated with three or more
TCR$\alpha$.

Convergent recombination of $\beta$ can create clones that
  shares their $\beta$ but not their $\alpha$ chains. This effect can
  be quantified using the generation and thymic selection model
  introduced in the previous paragraph. Simulations with the
  same sample sizes as the data show that such convergent
  recombination is predicted to happen with a rate of $0.5\%$, and
  thus cannot explain the data. However, there is another effect at
  play: cells divide around 5 times between $\beta$ and $\alpha$
  recombination, which leads to clones with the same $\beta$ chain but with up to $2^5\sim 30$ distinct $\alpha$ chains. A simulation considering these two effects together (see Methods) predicts a sharing fraction of $3\%$, consistent with the fraction observed empirically.

\section*{Discussion}

Analysing computationally reconstructed pairs of TCR $\alpha$ and $\beta$ chains, as well as $\alpha$-$\alpha$ and $\beta$-$\beta$ pairs, allowed us to quantify the various steps of sequence generation, rescue mechanisms, convergent selection, and sharing that were not accessible from just single-chain data.

Pairing $\alpha$ chains in single cells revealed correlations that were suggestive of a parallel and processive mechanism of VJ recombination in the two chromosomes. These signatures were well recapitulated by a simple computational model of successive rescue recombinations. Our model is similar to that of \cite{WarmflashModelTCRGene2006}, but differs in its details and parameters, as the original model could not reproduce the correlation pattern of the data. 

We estimated that $\sim 28\%$ of cells express two $\alpha$ chain, higher than a recent report of 14\% using single-cell sequencing \cite{Han2014}. However, this fraction is very hard to assess experimentally from high-throughput sequencing, as material loss can lead to its underestimation. While our estimate is indirect, we expected it to be more robust to such loss.

Our finding that the statistics of the two chains are largely independent of each other ---\,with only a weak correlation between $V_\beta$ and $(V_\alpha,J_\alpha)$ usage\,--- is in agreement with recent observations using direct single-cell chain pairing \cite{Grigaityte2017}. While independence between the $\alpha$ and $\beta$ recombination processes is perhaps expected because they occur at different stages of T-cell development, it is worth emphasizing that the absence of correlations reported here involves coding TCR$\alpha\beta$ sequences, which are believed to be largely restricted by thymic selection. This restriction can introduce correlations, notably through negative selection which could forbid certain $\alpha\beta$ combinations. Our results do not exclude such joint selection, but suggests that it does not introduce observable biases.
The independence between the two chains implies that the entropies of the two generation processes can be simply summed to obtain the entropy of the full TCR$\alpha\beta$. Taking the values previously reported in \cite{Mora2016e} of 26 bits for the $\alpha$ chain, and 38 bits for the $\beta$ chain, yields 64 bits for the TCR$\alpha\beta$, {\em i.e.} a diversity number of $2^{64}\approx 2\cdot 10^{19}$.

The independence between the chains also allowed us to make
predictions about the amount of TCR repertoire overlap one should
expect between samples from different individuals. Our analysis
predicts that sharing of $\alpha\beta$ pairs between two samples
should be rare, and that sharing between more than two is
exceptional. In a recent report \cite{Grigaityte2017}, 26
TCR$\alpha\beta$ pairs were found to be shared between any 2 of 5 individuals. Our result indicate that such a high level of sharing cannot be explained by convergent recombination alone: by simulating samples of the same size as in \cite{Grigaityte2017}, we estimated a total expected number of $0.001$ sequences between all their pairs (see Methods). The much higher number of shared sequences reported in the original study may result from over-correcting for sequencing errors, or alternatively from strong convergent selection in all 5 donors. A clonotype expansion of $10^4$ (not unexpected in the context of an immune response, see e.g. \cite{pogorelyy_precise_2018}) would be sufficient to explain this result.

Future studies collecting the $\alpha\beta$ repertoires of more individuals, as promised by the rapid development of single-cell sequencing techniques, will help us get a more detailed picture of the diversity and sharing properties of the TCR$\alpha\beta$ repertoires. Our analysis provides a useful baseline against which to compare and assess the results of these future works.

\section*{Methods}

\small{
\subsection*{Generation model}
The generation model was obtained and used through the IGoR software \cite{MarcouHighthroughputimmunerepertoire2018}. The \href{https://github.com/qmarcou/IGoR}{IGoR} software is able to learn, from out-of-frame receptor sequences, the statistics of a V(D)J recombination process. We don't use IGoR in its inference capacity here, but rather rely on the pre-inferred recombination model for TRA and TRB chains in humans supplied with IGoR, as the recombination process is widely shared between individuals \cite{Murugan2012}. Briefly, the probabilities of recombination of $\alpha$ and $\beta$ chains factorize as:
  \begin{equation}
    \begin{split}
    P_{\text{recomb}}^{\alpha}& = P\left(V, J\right) P\left(\text{del}V \middle| V\right) P\left(\text{del}J\middle|J\right) P\left(\text{ins}VJ\right) \\ & \times \prod_{i}^{\text{ins}VJ} P_{VJ}\left(n_i \middle| n_{i-1} \right),
  \end{split}
  \end{equation}
  \begin{equation}
    \begin{split}
    P_{\text{recomb}}^{\beta} &= P\left(V, D, J\right) P\left(\text{del}V \middle| V\right) P\left(\text{ins}VJ\right) \\ &\times P\left(\text{del}D5' \text{del}D3' \middle| D\right) P\left(\text{ins}DJ\right) \\ &\times P\left(\text{del}J\middle|J\right)  \prod_{i}^{\text{ins}VD} P_{VD}\left(m_i \middle| m_{i-1} \right) \\ & \times\prod_{i}^{\text{ins}DJ} P_{DJ}\left(r_i \middle| r_{i-1} \right),
  \end{split}
  \end{equation}
where $(n_i),(m_i),(r_i)$ are the inserted nucleotides at the VJ, VD, and DJ junctions. IGoR infers these probabilities through an Expectation-Maximization algorithm as described previously.
  
We rely on IGoR for:
\begin{itemize}
\item The generation of synthetic sequences with the same statistic as V(D)J recombination, which we use to predict sharing between individuals.
\item The computation of the probability of generation of a sequence $s$ by summing over all the scenarios that are compatible with it, $P_{\text{gen}}(s) = \sum_{\text{scenario}\to s} P_{\rm recomb}(\text{scenario})$, which allows us to generate Figure \ref{fig:sharing}. We also use this feature to predict sharing between very number large of sequences using Eq.~\ref{Mkm} (see \cite{ElhanatiPredictingspectrumTCR2018} for details).
\end{itemize}

\subsection*{Pairing of sequences}
We use the data and method of \cite{HowieHighthroughputpairingcell2015} to infer pairing from sequencing data of cells partitioned in $W=95$ wells (instead of 96 as erroneously reported in the original paper, as one of the wells did not provide any results). We calculate the p-value that two sequences each present in $w_1$ and $w_2$ well are found together in $w_{12}$ wells, under the null model that they are distributed randomly and independently:
$p(w_{1,2}, w_1, w_2, W) = \sum_{u \geq w_{12}}{{w_1 \choose u}{W - w_1 \choose w_2 - u}}/{{W \choose w_2}}$.

We first select all the pairs under a given p-value threshold ($10^{-4}$). For $\alpha-\alpha$ and $\beta-\beta$ pairs, we apply a threshold on their Levenshtein distances in order to remove most of the false pairings (pairing of near identical sequences due to sequencing errors). Then for each pair of well occupation numbers $(w_1, w_2)$, we set the p-value threshold so that false discovery rate (using the Benjamini–Hochberg procedure) is always less than $1\%$. Compared to the analysis of Ref.\,\cite{HowieHighthroughputpairingcell2015}, where the discreteness of the p-value distribution was taken into account by using a permutation algorithm, our approach is more conservative, as we worried about the potential effect of fake pairings on the false discovery rate. Thus our reported number of pairs (Table 1) slightly differs from that reported in the original study.

\subsection*{Information quantities}
The mutual information (in bits) of two variables $X, Y$ with a joint distribution $p(x,y)$ is defined by: $I(X, Y) = \sum_{x, y} p(x, y) \log_2[{p(x,
      y)}/{(p(x) p(y))}]$.
We estimated it from the empirical histogram of $(x,y)$ using a finite size correction \cite{SteuermutualinformationDetecting2002}, ${(n_X n_Y - n_X - n_Y + 1)}/{2 N \log(2)}$, where $N$ is the sample size, and $n_A$ is the number of different values the variable $A$ can take.

In the specific case of sequences in paired cells, a better correction can be obtained by computing the mutual information between shuffled sequences, where the two chains are assorted at random.

The Kullback-Leibler divergence between two distribution $p(x)$ and $q(x)$ of a variable $X$ is given by: $D_{\rm KL}(p\Vert q)=\sum_x p(x)\log_2(p(x)/q(x))$.

\subsection*{Simulation of the rescue process}
\label{sec:simu_rescue}
The V and J genes are indexed by $i$ and $j$ from most proximal to most distal along the chromosome: $V_i$, $i=1,\ldots L_V$ and $J_j$, $j=1,\ldots L_J$. In the first recombination attempt of the first chromosome, the model picks the V and J gene indices $i_1$ and $j_1$ from a truncated geometric distribution, $P(i_1=i)\propto(1-p)^{i-1}$ (and likewise for $j_1$), with $p=0.05$. The same process is simulated for the second chromosome.
With probability $2/3$ for each chromosome, the recombination fails. If both chromosome fail, a second recombination takes place on each between more distal genes indexed by $i_2>i_1$ and $j_2>j_1$, distributed as $P(i_2=i)\propto(1-p)^{i_2-i_1-1}$ (and likewise for $j_2$), to reflect observations that successive recombination occur on nearby genes in the germline \cite{PasqualQuantitativeQualitativeChanges2002}. If recombination repeatedly fail on both chromosomes, the process is repeated up to 5 times  \cite{MurphyJanewayImmunobiology9th2016}. This model is similar to that of \cite{WarmflashModelTCRGene2006}, where a uniform instead of a geometric distribution was used.

\subsection*{Bounds on rescue probabilities}
Non coding sequences can only appear in the TCR repertoire if they share a cell with a functional sequence. The probability of such a cell to appear in the selection process is $A=p_{\rm nc} (p_{\rm r} + p_{\rm r}') p_{\rm c} p_{\rm f}$. The probability for a cell to possess only one functional receptor is $B=p_{\rm c}p_{\rm f} (1 - p_{\rm r}')$, while the probability to possess two receptors and at least one functional one can be written as $C=p_{\rm c} p_{\rm f} \left[ p_{\rm r} \left(1 - p_{\rm c} p_{\rm f} \right) + p_{\rm r}' \right]$. The proportion of non-coding reads is thus $A/(B+2C)$, which gives Eq.~\ref{eq:proportion_beta}.

\subsection*{Simple model of selection based on the V genes segments}

We have shown that the pairs $V_{\alpha}-V_{\beta}$ and $J_{\alpha}-V_{\beta}$ were not independent (Fig.\,2). In this section we define the simplest model that can reproduce these correlations.
The marginal distributions $p_{V_\alpha, J_{\alpha}}$ and $p_{V_\beta}$, coupled with the experimental pairing data can be used to obtain selection factors $q_{V_\alpha, J_{\alpha} V_\beta}$:
\begin{equation}
  p(V_{\alpha}, J_{\alpha}, V_{\beta}) = p_{V_\alpha, J_{\alpha}}\  p_{V_\beta} \  q_{V_{\alpha}, J_{\alpha}, V_{\beta}}
\end{equation}
By adding a tunable temperature, we can modify the level of selection we want to observe:
\begin{equation}
  p(V_{\alpha}, J_\alpha, V_{\beta}) \propto p_{V_\alpha} \ p_{V_\beta} \  \left(q_{V_{\alpha}, V_{\beta}}\right)^{1/T}
\end{equation}
When $T \rightarrow 0$, the selection conserves only a few specific pairs of V, while for $T \rightarrow \infty$ there is no selection. This modifies the mutual information between $V_\alpha$ and $V_\beta$ in the same cell, but also, because $V$ and $J$ on the same chromosome are not independent, the mutual information between $V_\alpha$ and $J_\beta$. In Fig.\,S11, we show the evolution of the mutual information between $V_\alpha$, $J_{\alpha}$, $V_\beta$ and $J_\beta$ as a function of $T$. The model underestimates the mutual information between $V_{\alpha}$ and $J_{\beta}$ which hints that it may be necessary to also include $J_{\beta}$ in the selection model. 

\subsection*{Copy number distributions}
We fit  the empirical distribution of reads per coding chain, $\rho_{\text{c}}$, with a mixture of two distributions (Fig.\,S4): $\rho_{\text{e}}$, corresponding to chain sequences that could be paired with a non-coding sequence of the same type and thus believed to be expressed; and $\rho_{\text{nc}}$ corresponding to non-expressed sequences and learned from non-coding sequences. The fit is done by  mean square error minimization: $\int \mathrm{d}x (\rho_{\text{c}}(x) - \lambda_1 \rho_{\text{nc}}(x) - \lambda_2 \rho_\text{e}(x) )^2$. The fraction of expressed chains among coding ones is then given by $p_{\rm e}=\lambda_2/(\lambda_1+\lambda_2)$. Calling $p_{2\alpha}$ the proportion of cells with two expressed $\alpha$, the resulting fraction $p_{\rm e}$ of $\alpha$ sequences that are expressed should be $p_{\rm e}=(2p_{2\alpha}+(1-p_{2\alpha}))/2$, hence $p_{2\alpha}=2p_e-1$.

We find a value $p_{e}^{\alpha} = 28 \% \pm 10 \%$, not compatible with the value of $14 \% \pm 3\%$ obtained in \cite{Han2014} (19 out of 139 cells in which at least one productive sequence was found). But the authors of \cite{Han2014} make their estimate by sequencing cDNA, which can lead to different drop-out rates depending on the nature of the sequence. Silenced productive sequences or non-productive sequences are less expressed and their drop-out rates are higher. They find two TCRA (productive or not) in only $58\%$ of cells, while both TCRA are expected to recombine \cite{NiederbergerAllelicExclusionTCR2003}. In this context the $14\%$ rate can only be understood as a lower bound. Assuming that non-productive and silenced sequences are expressed in similar quantities, we obtain an estimate for $p_{e}^{\alpha}$ of $24\% \pm 5\%$ (19 out of the 80 cells which had two sequences, productive or not) from their data, which is consistent with our result.

\subsection*{Sharing estimation}
\label{sec:method_sharing}

We follow the methods of \cite{ElhanatiPredictingspectrumTCR2018}. A large number of productive  $\alpha$ and $\beta$ chain pair sequences are generated through a stochastic model of recombination using IGoR \cite{MarcouHighthroughputimmunerepertoire2018}. Each TCR$\alpha\beta$ amino-acid sequence is then kept if its normalized hash (a hash is a deterministic but maximally disordered function) is $\leq q=q_\alpha q_\beta$, so that a random fraction $q$ of sequences passes selection.
The values of $q_\alpha$ and $q_\beta$ are learned from rarefaction curves showing the number of unique amino-acid sequences of each chain as a function of the number of unique nucleotide sequences (Fig.\,S5), using the analytical expressions given in \cite{ElhanatiPredictingspectrumTCR2018}.

The predictions for the number of shared TCR$\alpha\beta$ nucleotide sequences reported in Fig.\,\ref{fig:sharing}B, as well as the estimation of the sharing between the full repertoire of two individuals, are computed using the analytical expressions of \cite{ElhanatiPredictingspectrumTCR2018}.
If $N$ sequences are sampled in $m$ individuals, the expected number of sequences which will be found in exactly $k$ individuals is:
\begin{equation}\label{Mkm}
  M_{k, m}(N) = \int_{0}^{\infty} \mathrm{d}p P(p) {{m}\choose{k}} e^{N p (m -k)} \left(1 - e^{-N p}\right)^{k} 
\end{equation}
Without selection $P(p)$ is the probability density function for of sequences probabilities. We used this formula with $p=P_{\rm gen}/q$ for selected sequences, and $p=0$ otherwise. The integral in Eq.~\ref{Mkm} is evaluated using a Monte Carlo simulation.
Derivations and details about the Monte Carlo simulation can be found in \cite{ElhanatiPredictingspectrumTCR2018}. We use this formula to estimate the proportion of full receptors shared between two individuals.

\subsection*{$\beta$ sharing}
\label{sec:beta_sharing}
The results of \cite{ElhanatiPredictingspectrumTCR2018} can also be used to estimate the theoretical proportion of clonotypes sharing a $\beta$ in a sample of size $N$. This sharing is due to two phenomena: the possibility of generating twice the same $\beta$ sequence and the division stage between the recombination of $\beta$ and $\alpha$. To simulate the first mechanism we can, following \cite{ElhanatiPredictingspectrumTCR2018}, generate an important number of $\beta$ sequences (in-frame, no-stop codons) with IGoR, associate to each of them a hash between $0$ and $1$ and then only keep the sequences whose hash is lower than $q_\beta$ to simulate the selection. The cellular division between $\beta$ and $\alpha$ recombination creates $30$ cells with the same $\beta$ and different $\alpha$. Some of these cells won't have a functional $\alpha$ receptors, while others will not pass selection, while there is no precise way to quantify how many cells survive, we can consider an estimate of roughly $n_d \approx 10$ cells. Because the probability $p(s)$ of generating a given sequence is so low, this increase in cell number multiplies $p(s)$ by $n_d$, hence corresponds to a change $q_\beta \rightarrow q_\beta/n_d$. Then, for $10^5$ sequences and $n_d = 10$, we find that $\approx 3\%$ of clonotypes are expected to share their $\beta$ sequence with another TCR.

\subsection*{Data and code availability}
All the code and
curated data used to produce the analyses of this paper are available at
\url{https://github.com/Thopic/TCR_pairings}.

}

\bibliographystyle{pnas}

\onecolumngrid

\clearpage

\appendix

\begin{table}[!ht]
\centering
\caption{{\bf Number of $\alpha$-$\beta$, $\alpha$-$\alpha$ and $\beta$-$\beta$ statistically significant pairs in each of the three experiments from \cite{HowieHighthroughputpairingcell2015}.} Samples were obtained from two human subjects $X$ and $Y$ and divided in three experiments (experiment $1$, $2$, and $3$), with different sequencing depths and different subjects: experiment $3$ contains only sequences from $X$, while experiments $1$ and $2$ contain sequences from both subjects.
}

\begin{tabular}{|c|c|c|c|c|c|c|}
  \hline
Exp. & \# cells & unique $\alpha$ & unique $\beta$ &   pairs $(\alpha, \beta)$ & pairs $(\alpha, \alpha)$ & pairs $(\beta, \beta)$ \\
\hline
1 & $3.8 \times 10^5$   & $1.8\times 10^6$  & $1.7 \times 10^6$  & 1098 & 336 & 30  \\
2 & $1.5 \times 10^7$ & $2.7 \times 10^7$ & $3.3 \times 10^7$  & 79420 & 47665 & 7795  \\
3 & $1.5 \times 10^7$ &  $5.1 \times 10^7$ & $6.3 \times 10^7$  & 129757 & 89957 & 15361 \\
\hline
\end{tabular}

\label{tab:number_pairs}
\end{table}

\begin{table}[!ht]
  \centering
  \caption{{\bf Length distribution and Kullback-Leibler (KL) divergence from the unselected (non-coding) ensemble for different types of sequences: functional (and expressed), coding, and non-coding. The error on the standard deviation of the length (estimated by bootstrap) is always lower than 0.2.}}
  \begin{tabular}{|c||p{1.7cm}|p{1.7cm}|p{1.7cm}||c|p{2cm}|p{2cm}|}
      \hline
  {chain} & \multicolumn{3}{c||}{length: mean $\pm$ st. deviation (nt)}& {Gene} & \multicolumn{2}{p{4cm}|}{\centering KL divergence (bits)}  \\
[.5em]
&  functional& coding & non-coding & & functional &   coding  \\
  \hline
\multirow{2}{*}{$\alpha$}  & \multirow{2}{*}{$42.0 \pm 5.00$} & \multirow{2}{*}{$39.12 \pm 6.67$} & \multirow{2}{*}{$40.0 \pm 7.00$} & $V_\alpha$ & $0.66 \pm 0.05$ & $1.39 \pm 0.01$  \\
   &  & & & $J_\alpha$ & $0.110 \pm 0.005$  & $0.119 \pm 0.004$  \\
\multirow{2}{*}{$\beta$}  & \multirow{2}{*}{$44.1 \pm 5.03$}& \multirow{2}{*}{$43.17 \pm 6.22$}  & \multirow{2}{*}{$43.4 \pm 7.82$} & $V_\beta$ & $1.09 \pm 0.06$  & $1.03 \pm 0.18$   \\
 &   & & & $J_\beta$ & $0.12 \pm 0.004$ & $0.051 \pm 0.008$  \\
\hline
\end{tabular}
\label{tab:shannon_entropy}
\end{table}

\setcounter{table}{0}
\renewcommand{\thetable}{S\arabic{table}}%
\setcounter{figure}{0}
\renewcommand{\thefigure}{S\arabic{figure}}%

\begin{figure*}
  \begin{center}
\includegraphics[width=\linewidth]{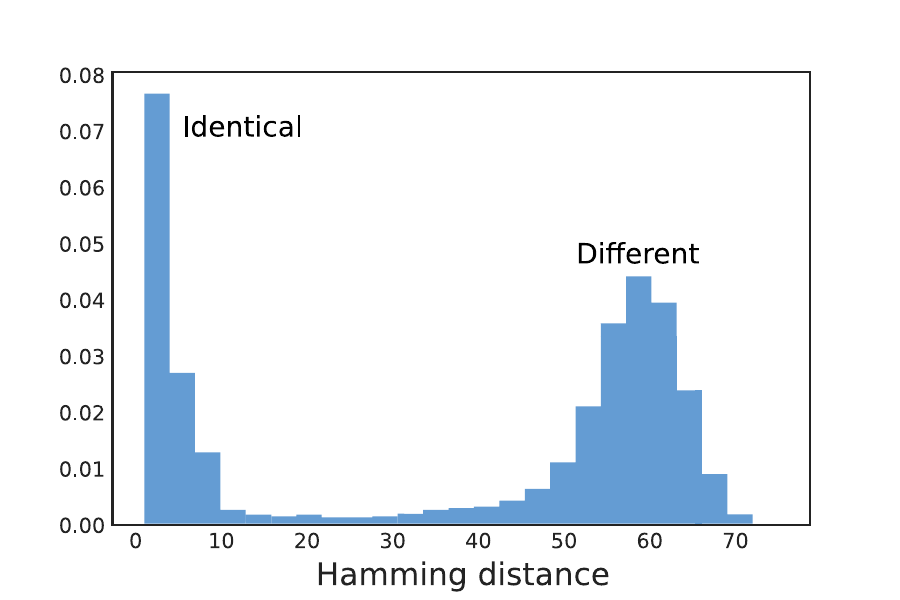}
\caption{{\bf Hamming distance between two TCR$\beta$ sequences identified as paired.}  Near-identical paired sequences are in their vast majority due to sequencing error. The Hamming distance permits to separate effectively these sequences from actually different sequences extracted from the same clone. A similar behaviour is observed for TCR$\alpha$ chains.
}
\end{center}
\end{figure*}

\begin{figure*}
  \begin{center}
\includegraphics[width=\linewidth]{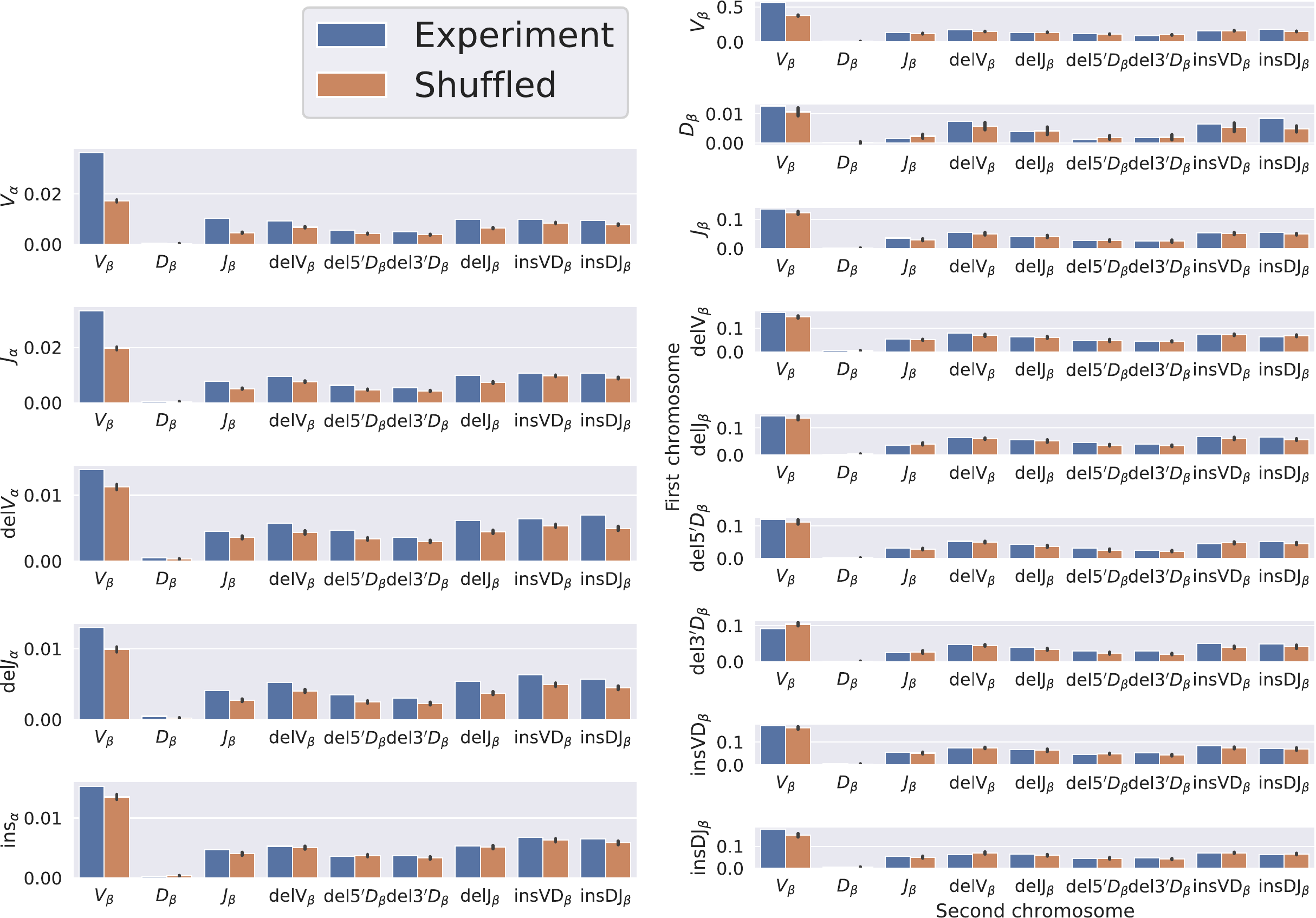}
\caption{{\bf Comparison between the observed mutual information and the null for $\alpha$-$\beta$ (A) and $\beta$-$\beta$ pairs (B).} The null distribution is obtained by shuffling the pairs, the error-bar represents the standard deviation over multiple shuffling. We consider the raw mutual information, not corrected with the shuffled distribution, contrary to Fig.\,2. With a false discovery rate of 0.01 (using the Benjamini–Hochberg procedure) and assuming a Gaussian distribution for the mutual information of shuffled sequences, we find that, for $\beta-\beta$ pairings, the only pairs of features passing the test are (in order of significance) $V_1-V_2$, $V_1-\text{Ins}DJ_2$ and $\text{Del}3^\prime D_1-\text{Ins}DV_2$. By contrast, for $\alpha-\beta$ pairing, with the same false discovery rate (0.01), 36 out of the 45 possible feature pairings are significant.
}
\end{center}
\end{figure*}

\begin{figure*}
  \begin{center}
\includegraphics[width=\linewidth]{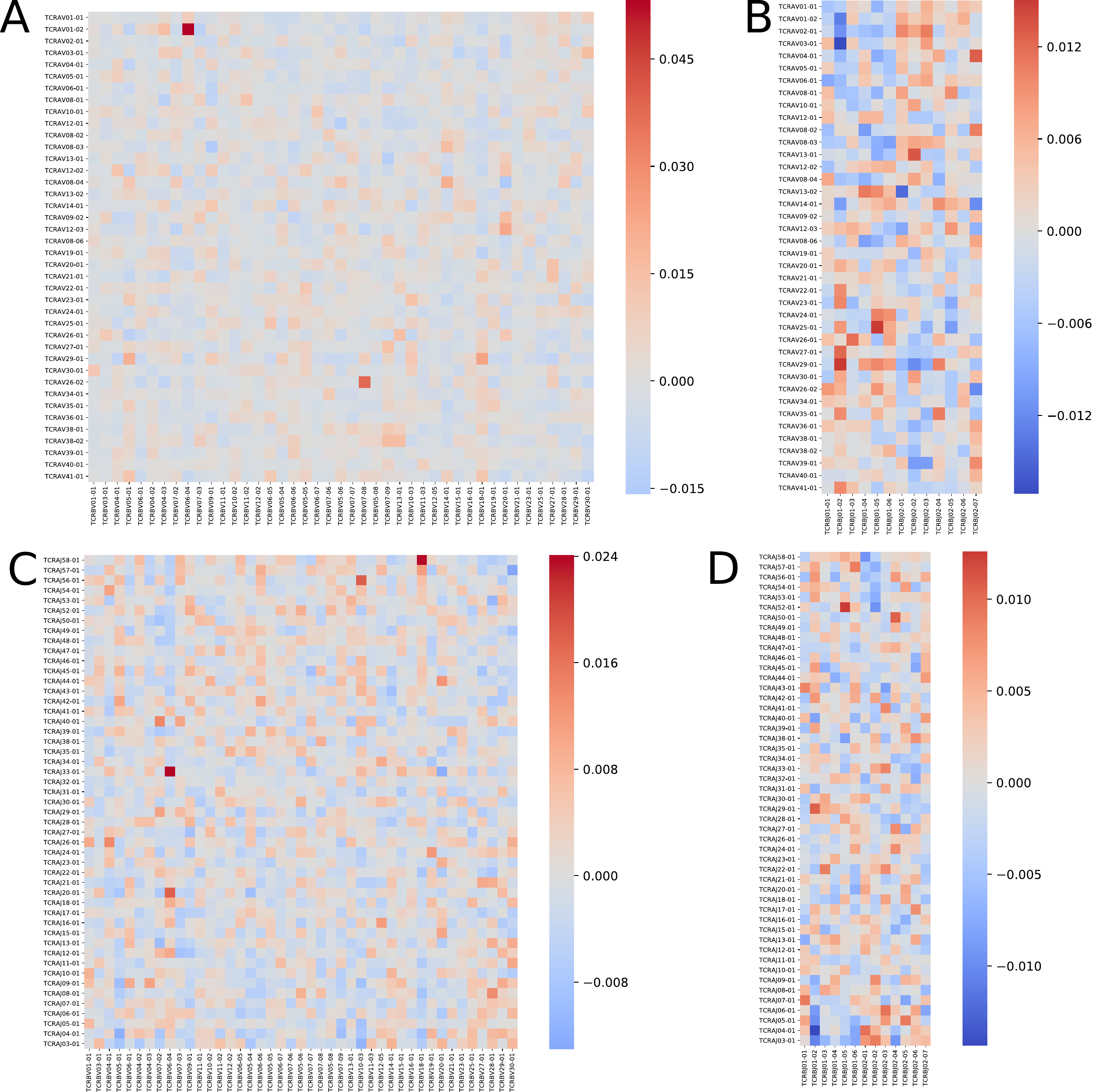}
\caption{{\bf Pearson correlation coefficient between TCRA and TCRB genes. $V_{\alpha}$ - $V_{\beta}$ (A), $V_{\alpha}$ - $J_{\beta}$ (B), $J_{\alpha}$ - $V_{\beta}$ (C) and $J_{\alpha}$ - $J_{\beta}$ (D).} The correlation are generically small and do not show a particular structure.
}
\end{center}
\end{figure*}

\begin{figure*}
  \begin{center}
\includegraphics[width=\linewidth]{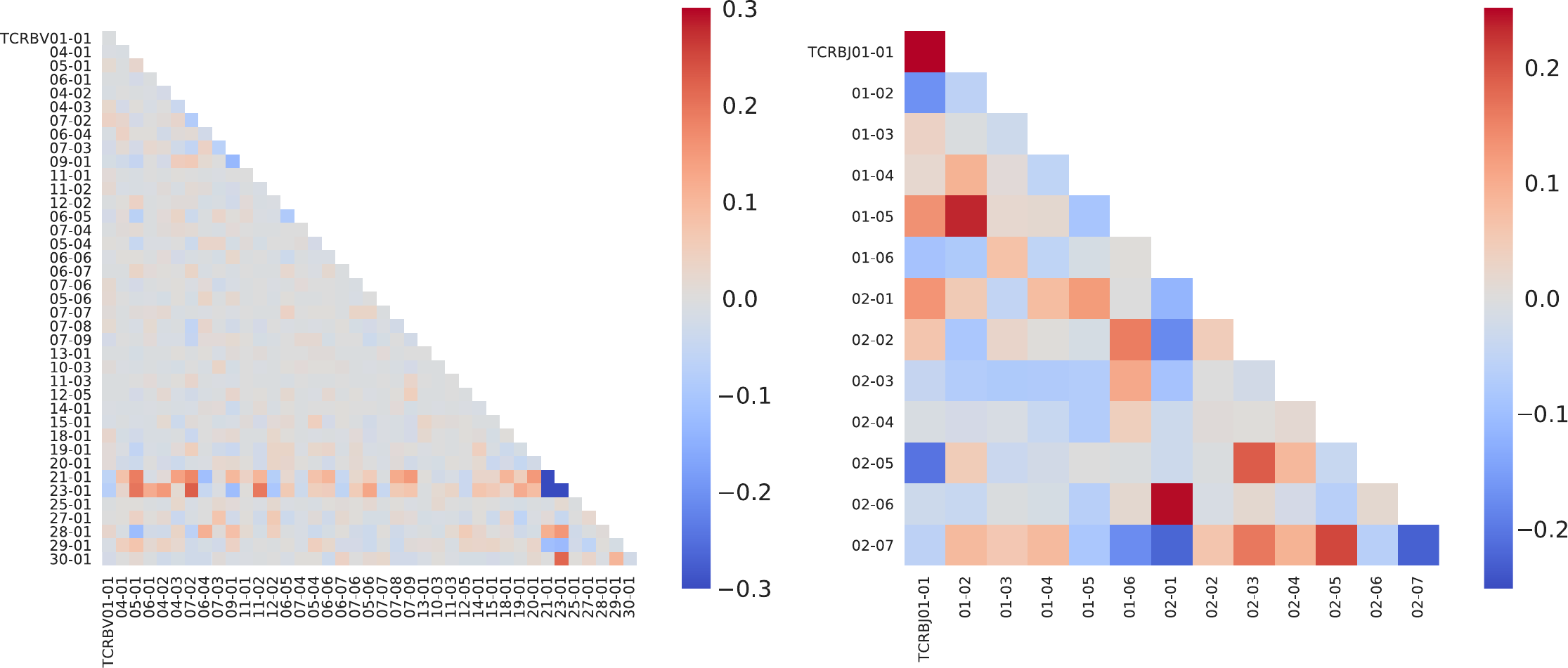}
\caption{{\bf Normalized covariance between V (left) and J (right) gene usages of pairs of $\beta$ sequences found in the same clone.} The V21-01 and V23-01 genes are non-functional pseudogenes and are thus anticorrelated.
}
\end{center}
\end{figure*}

\begin{figure*}
  \begin{center}
\includegraphics[width=\linewidth]{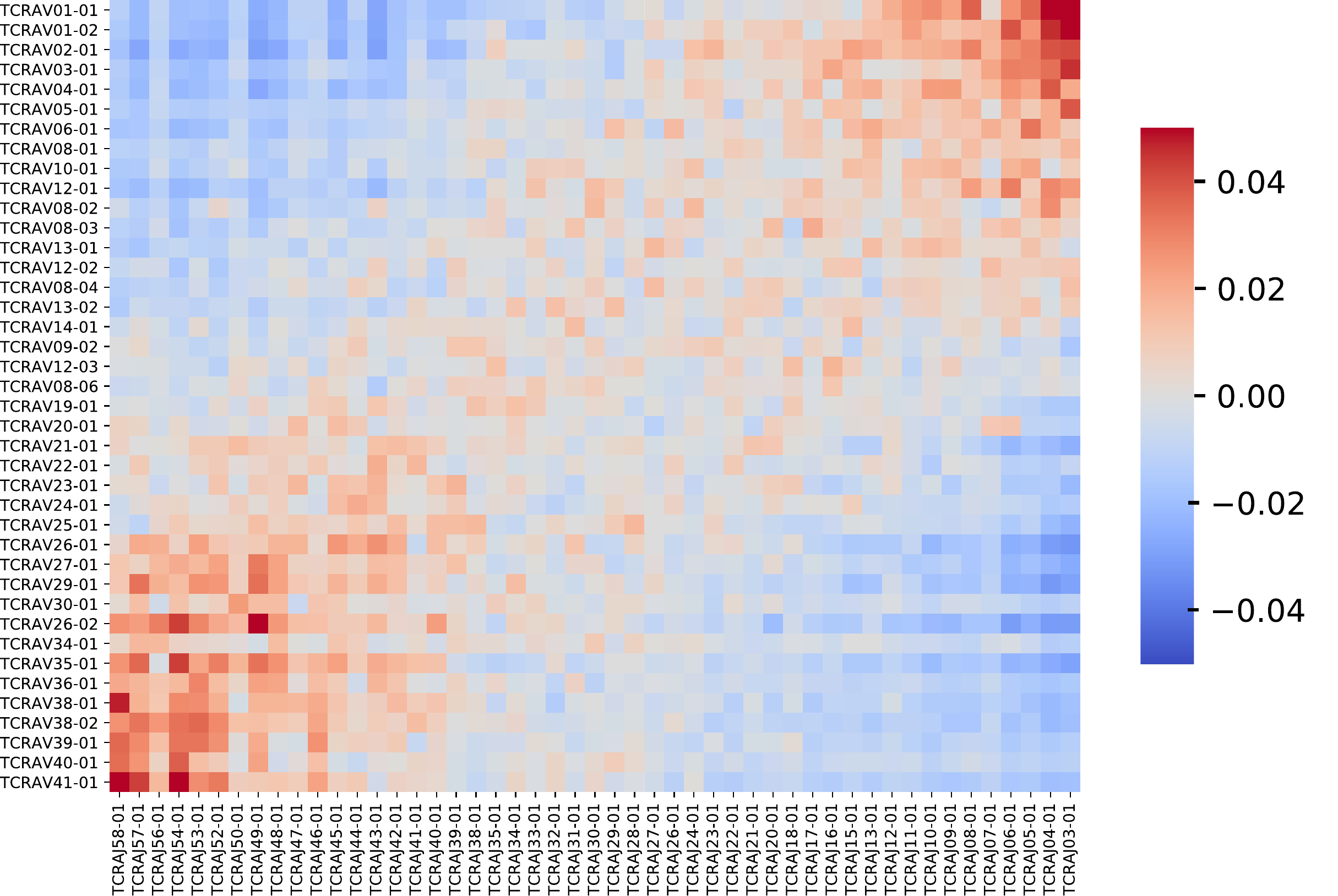}
\caption{{\bf Pearson correlation between the $V_{\alpha}$ gene fragment on the first chromosome and the $J_{\alpha}$ gene fragment on the second chromosome.} The correlations observed in Fig 3A and 3B are also observed here.
}
\end{center}
\end{figure*}

\begin{figure*}
  \begin{center}
\includegraphics[width=\linewidth]{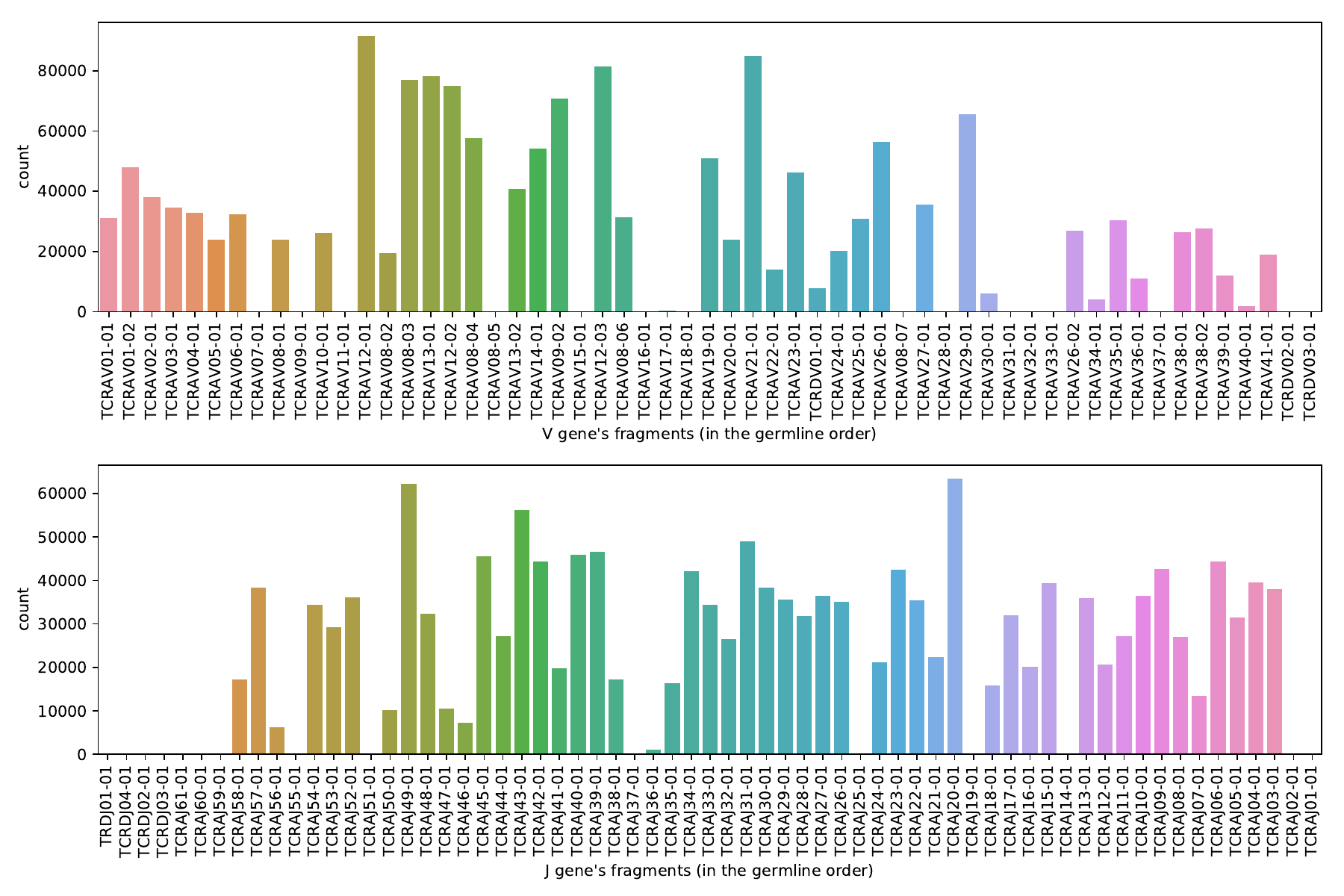}
\caption{{\bf Distribution of the V and J gene fragments. In both case, they are ordered along the germline, 5' to 3'.}
}
\end{center}
\end{figure*}

\begin{figure*}
  \begin{center}
\includegraphics[width=\linewidth]{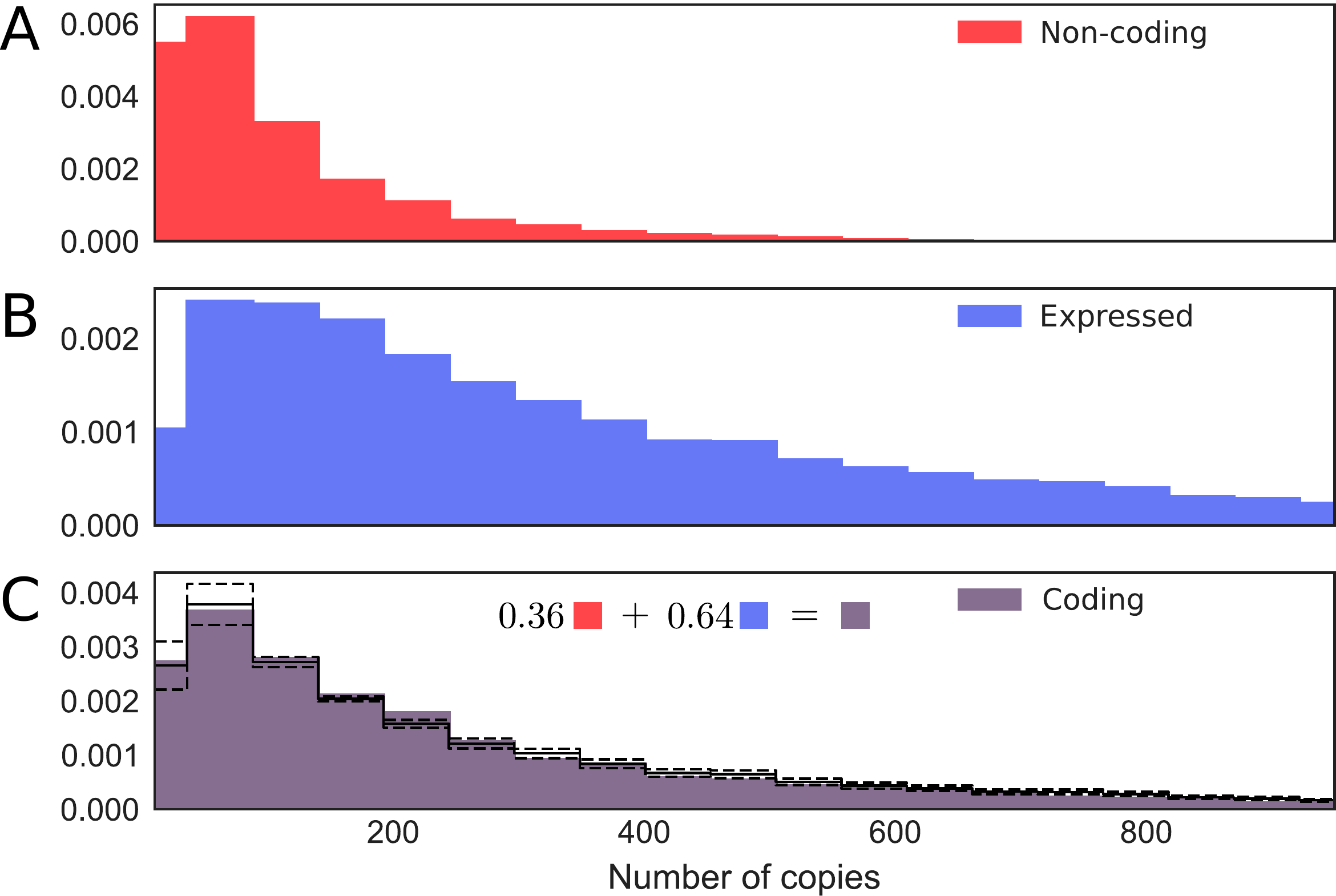}
\caption{{\bf Distribution of the number of reads of different types of TCR$\alpha$ RNA sequences.} (A) non-coding; (B) functional and expressed (i.e. paired with a non-coding sequence); (C) `just coding' sequences. In panel (C), the full line represents the best fit for a mixture of 36\% of non-coding and 64\% of functional, expressed sequences. Dashed lines show $10\%$ error intervals. To avoid biases in the comparisons, all sequences used in these distributions were paired.
}
\end{center}
\end{figure*}

\begin{figure*}
  \begin{center}
\includegraphics[width=\linewidth]{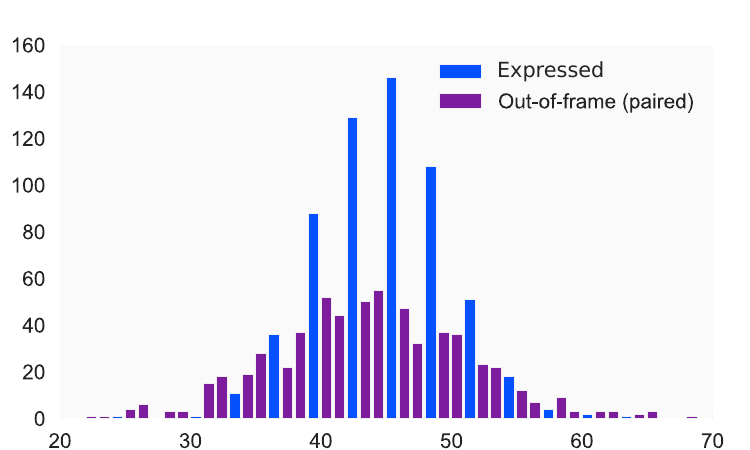}
\caption{{\bf CDR3 length distribution of expressed and out-of-frame TCR$\alpha$ sequences.} Expressed sequences have a narrowed distribution than unselected ones. All sequences used in these distributions were paired.
}
\end{center}
\end{figure*}

\begin{figure*}
  \begin{center}
\includegraphics[width=\linewidth]{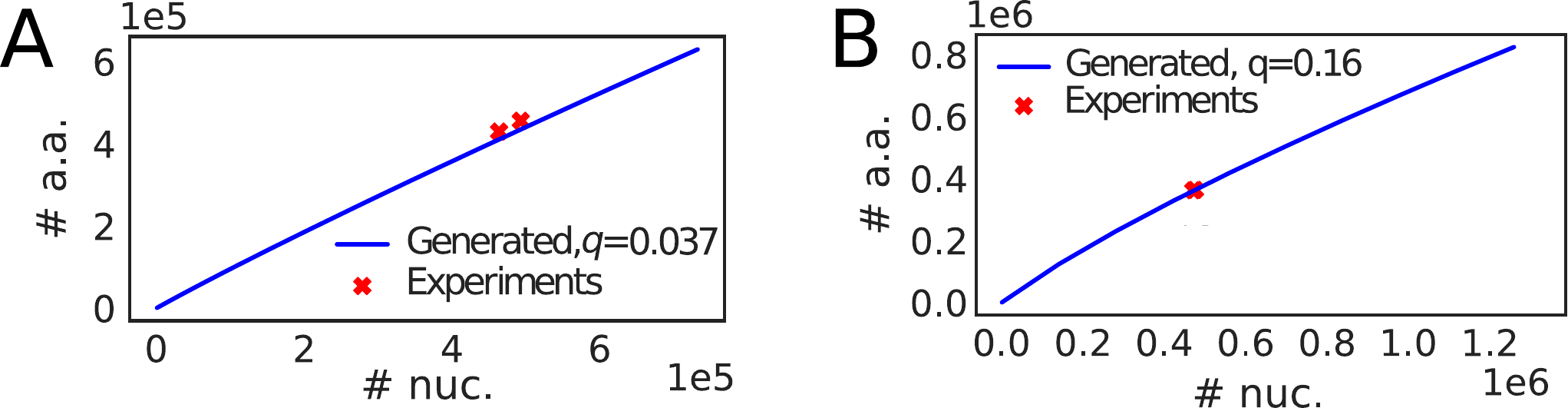}
\caption{{\bf Number of unique amino-acid (translated) sequences as a function of the number of  unique nucleotide sequences for (A) $\alpha$ and (B) $\beta$ chains.} Red crosses are experimental data, blue line comes  from simulations of the recombination model with random selection. For $\alpha$ the value of $q$ is inferred by least-square minimisation to be $q_\alpha=0.16$, while for $\beta$ we used the value of $q_\beta=0.037$ reported in Elhanati et al., {\em Immunological Reviews}, in press (2018).
}
\end{center}
\end{figure*}

\begin{figure*}
  \begin{center}
\includegraphics[width=\linewidth]{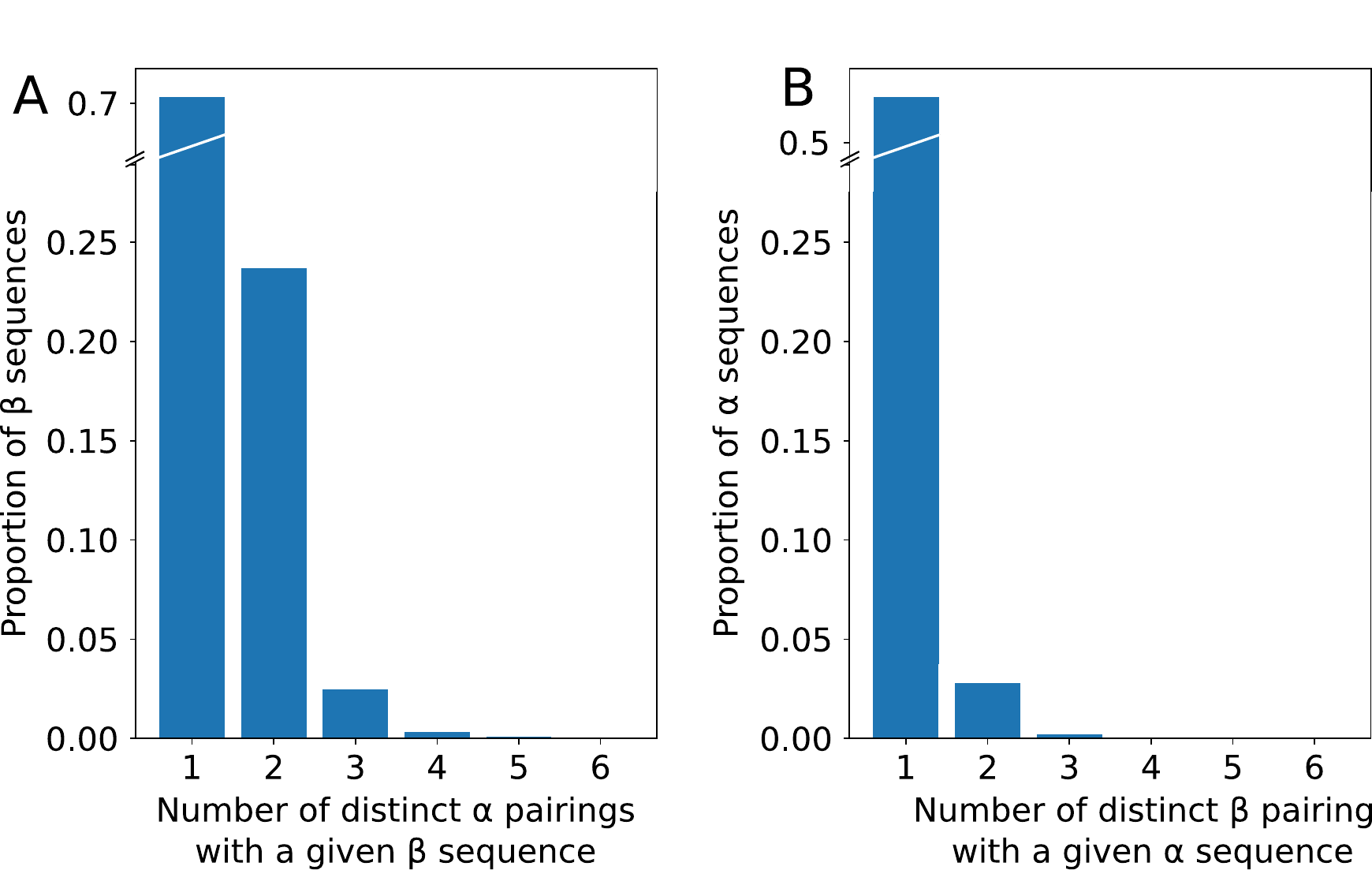}
\caption{{\bf (A) Distribution of the number of distinct $\alpha$ sequences that could be paired with a given $\beta$ sequence. (B) Distribution of the number of distinct $\beta$ sequences that could be paired with a given $\alpha$ sequence.} Only sequences that appear in at least a pairing are considered. Since sequences may be paired with 2 chains of the other type in a single cell, only chains with 3 or more associations unambiguously correspond to the convergent selection of that chain in different clones.
}
\end{center}
\end{figure*}

\begin{figure*}
  \begin{center}
\includegraphics[width=\linewidth]{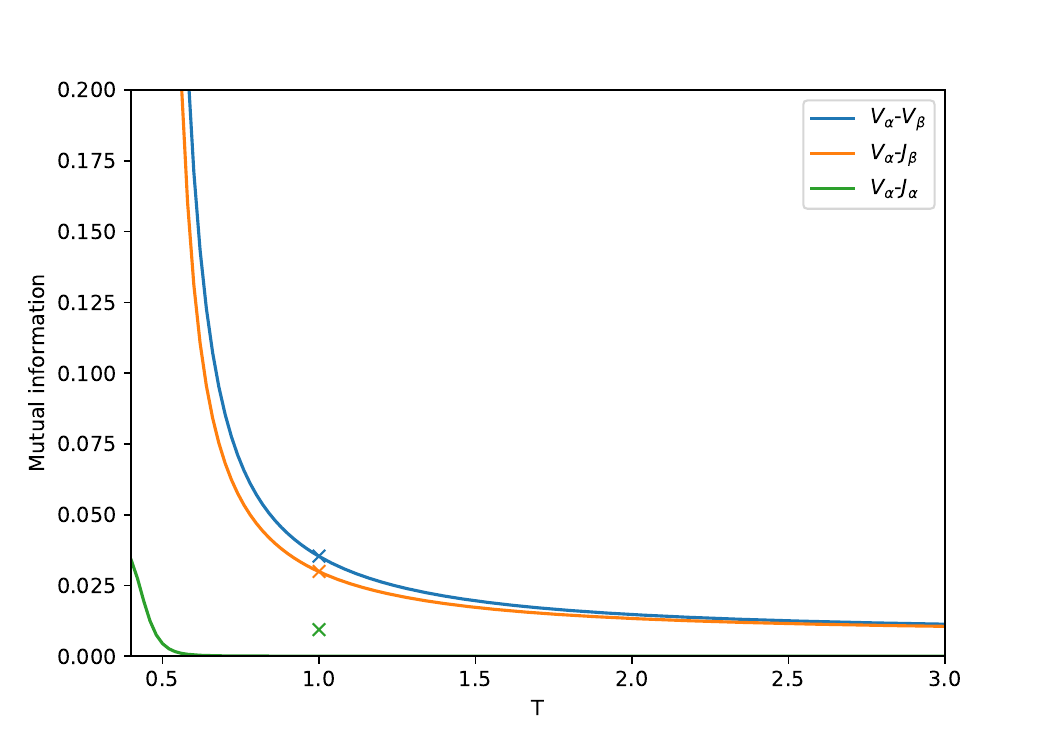}
\caption{{\bf The full blue (resp. yellow, green) line represent the mutual information between $V_\alpha$/$V_\beta$ (resp. $V_\beta$/$J_\alpha$, $J_\beta$/$V_\alpha$), as a function of temperature $T$, as described in the Methods section. The dot are the observed values in the dataset.}
}
\end{center}
\end{figure*}

\end{document}